\documentclass[lettersize,journal]{IEEEtran}

\usepackage{amsmath,amsfonts}
\usepackage{algorithmic}
\usepackage{algorithm}
\usepackage{array}
\usepackage[caption=false,font=normalsize,labelfont=sf,textfont=sf]{subfig}
\usepackage{textcomp}
\usepackage{stfloats}
\usepackage{url}
\usepackage{tabularx}
\usepackage{verbatim}
\usepackage{booktabs}
\usepackage{graphicx}
\usepackage{geometry}
\usepackage{svg}

\usepackage{cite}
\usepackage{enumitem} 
\begin{document}

\title{Positioning Error Compensation by Channel Knowledge Map in UAV Communication Missions}

\author{Chiya Zhang*, Ting Wang*, Chunlong He
\thanks{This work was supported by the National Natural Science Foundation of China under Grant 62101161,  62394294, and U20A20156. And this work was also supported by Foundation of National Key Laboratory of Radar Signal Processing  under Grant JKW202303.}
\thanks{*These authors contributed equally to this work. (email: 24B352015@stu.hit.edu.cn).}
\thanks{C. Zhang and T. Wang are with School of Electronic and Information Engineering, Harbin Institute of Technology, Shenzhen, 518055, China.
C. He is with Guangdong Key Laboratory of Intelligent Information
Processing, Shenzhen University, Shenzhen, 518060, China.}}


\maketitle

\begin{abstract}
    When Unmanned Aerial Vehicles (UAVs) perform high-precision communication tasks, such as searching for users and providing emergency coverage, positioning errors between base stations and users make it challenging to deploy trajectory planning algorithms. To address these challenges caused by position errors, a framework was proposed to compensate it by Channel Knowledge Map (CKM), which stores channel state information (CSI). By taking the positions with errors as input, the generated CKM could give a prediction of signal attenuation which is close to true positions. Based on that, the predictions are utilized to calculate the received power and a PPO-based algorithm is applied to optimize the compensation. After training, the framework is able to find a strategy that minimize the flight time under communication constraints and positioning error. Besides, the confidence interval is calculated to assist the allocation of power and the update of CKM is studied to adapt to the dynamic environment. Simulation results show the robustness of CKM to positioning error and environmental changes, and the superiority of CKM-assisted UAV communication design.
\end{abstract}


\section{Introduction}
\IEEEPARstart{U}{nmanned} Aerial Vehicles(UAVs) are increasingly applied in wireless communication systems due to their flexibility, low cost and quick deployment. When stationary base stations have limited coverage, particularly if they are destroyed, causing weak or nonexistent signal propagation, UAVs can quickly establish line-of-sight (LoS) propagation as substitutes for base stations. Consequently, the design of UAV trajectories, which involves optimizing their flight paths to maximize communication efficiency and reliability or minimize flight time, becomes a crucial topic of research and development in the field of wireless communications\cite{ref1}. 
\par
On the other hand, the channel model between UAVs and ground users is a critical aspect of ensuring reliable and efficient communication in UAV-IoT networks. Typically, this model is characterized by a combination of LoS and NLoS conditions\cite{ref2}.
A common UAV-to-ground (U2G) communication model involves large-scale path loss, shadowing effects, and small-scale multipath fading. For instance, the probability of having a LoS path is a key element and can be calculated based on the elevation angle and distance between the UAV and the ground user. 
Advanced models also integrate three-dimensional (3D) building footprints and terrain data to more accurately simulate urban environments, accounting for blockages and reflections that affect signal propagation\cite{ref3,ref4}. These models can be enhanced using machine learning techniques to predict channel characteristics dynamically, adapting to real-time environmental changes and mobility patterns.
Overall, accurate U2G channel modeling is essential for optimizing UAV deployment, improving coverage, and ensuring robust communication links in various UAV-assisted applications.
However, in a bad environment, factors such as obstacle reflection and propagation delay will amplify the positioning error, thereby affecting the channel model's estimation and furthermore influencing the applications' performance.
Therefore, the influences of positioning errors should be carefully considered during the application of the above models.

 In studies of UAV communication design, channel models are variably designed according to the different emphasis of the articles. The authors in \cite{ref5} used a simple sigmoid function to model the LoS probability. Based on that, \cite{ref6} investigated the coverage probability, average ergodic rate and area spectral efficiency based on a more realistic channel model, with height dependent LoS probabilities. Their results indicate that the long tail of the LoS probability function significantly impacts performance in sparse networks, where the base station (BS) density is low. Assumed under a static scence, the above models can not perform much well in dynamic environment. Since an imperfect CSI is more common, many work studied how to find a solution with imperfect CSI. In \cite{ref7}, the UAV-mounted base station (BS) first obtains the angle-of-arrivals (AoAs) information of the uplink signals from mobile users, thus enabling the estimation of the location information and CSI. Then, based on the estimated location information and CSI, a hybrid receiving beamforming and trajectory optimization problem has been formulated to maximize the weighted average throughput (WAT). The authors in \cite{ref8} considered a practical 3-D urban environment with imperfect CSI and sets an additional information, for example, the merged pheromone, to represent the state information of the UAV and environment as a reference of reward which facilitates the algorithm design.
\par

\begin{figure*}[t]
    \centering
    \includegraphics[width=0.8\linewidth]{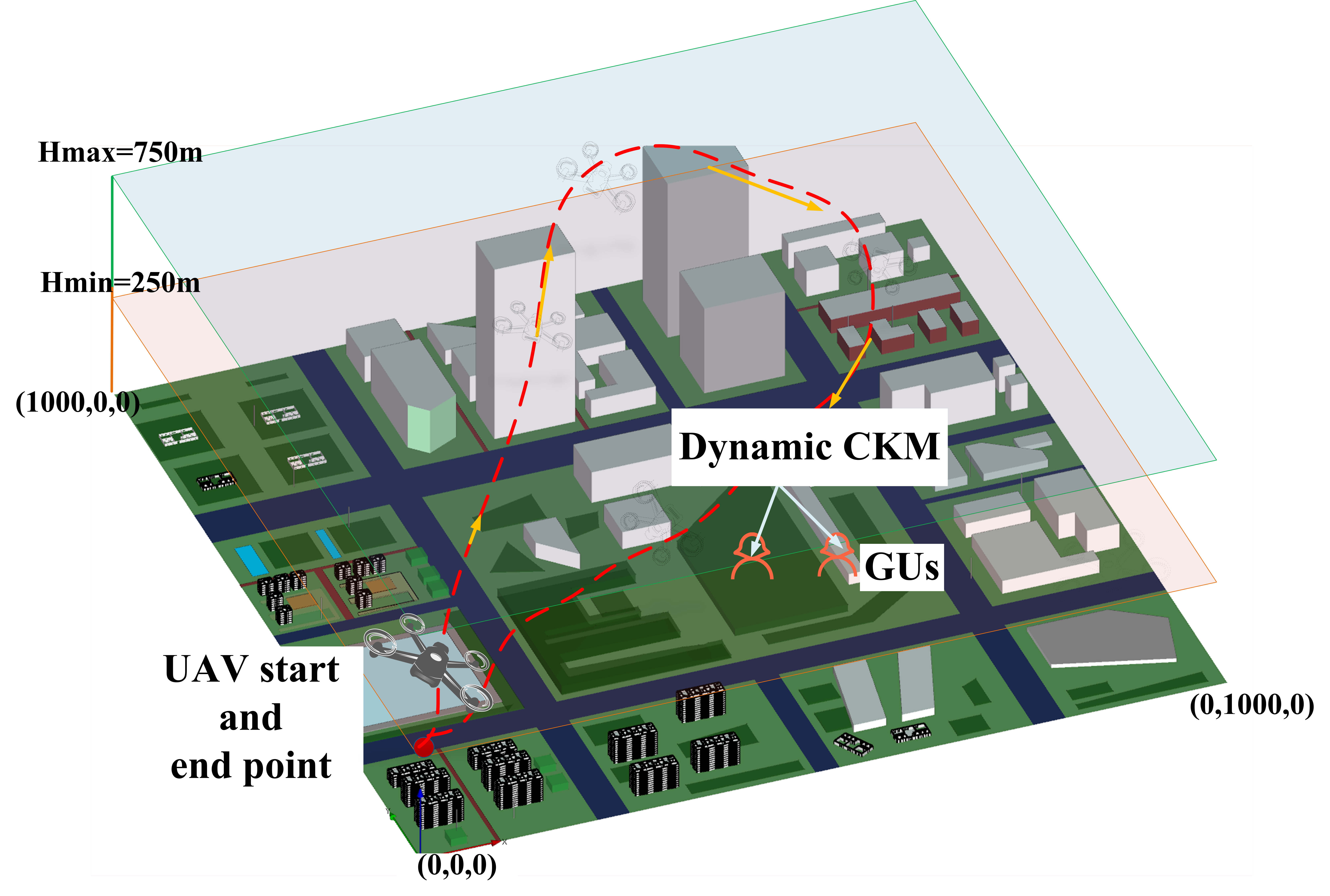}
    \caption{UAV-assisted Communication Scenario. Length is 1000m, width is 1000m, height is 750m, UAV flies between 250-750m,\\ 15 GUs are distributed randomly between 0-250m, and buildings and plants are randomly set under 250m}
    \label{fig2}
  \end{figure*}
 To provide a detailed spatial representation of radio signal strength and quality within a specific area, radio map has been proposed. The authors in \cite{ref9} optimized the accuracy of the radio map using a geostatistical tool named Kriging interpolation in cognitive radio networks. What's more, \cite{ref10} considered the distribution of the city and make fusion of radio and city information.
To better describe the intrinsic characteristics of the environment, the concept of Channel Knowledge Map(CKM) was proposed in \cite{ref11}. CKM is a site-specific database tagged with transceiver location pairs, which could give the real-time CSI acquisition and then used to design UAV trajectory by turning uncertain information into certain ones. 
\par
UAVs can get access to ground users easily, offering a high probability of LoS to offload data stably. Various work have investigated the trajectory design of UAVs under a communication scenario to get an optimization, like energy efficiency\cite{ref12},\cite{ref13},flight time\cite{ref14}-\cite{ref16}, age-of-information(AOI)\cite{ref17}, and throughput\cite{ref18}-\cite{ref20}. Specifically, \cite{ref12} proposed an offline-based online adaptive design to adapt to real-time wind based on the wind statistics. In \cite{ref13}, the joint problem of IRS phase shift, UAV trajectory and power allocation in the system was investigated, aiming to maximize the energy efficient. 
In some cases, flight time is a more important factor.
\cite{ref14} dispatched an UAV to assist data collection from multiple sensor nodes (SNs). They provided a completion time minimization design via jointly deciding the UAV trajectory and the SN assignment scheme. \cite{ref15} formulated the mission completion time under energy constraints to an optimization problem and decomposed it into three sub-problems: mission allocation, collecting time minimization and joint optimization of the former two.
AOI measures the freshness of information being delivered. \cite{ref16} fed the whole UAV-IoT system into an encoder network and outputted the visiting order of ground clusters. In bad environments such as earthquakes, optimization of AOI seriously affects the system performance.
To maximize the achievable sum rate of all IoT nodes, \cite{ref18} maximized the minimal UAV data collection throughput from GNs for both orthogonal multiple access (OMA) and non-orthogonal multiple access (NOMA) transmission, subject to the energy budgets at both the UAV and GNs. Authors in \cite{ref19} thought severe transmission delay unfairness within the network might appear, so they proposed a novel two-phase scheme to balance transmission delay of users and network fairness. \cite{ref20} proposed an energy-efficient cooperative relaying scheme to minimize the maximum (min-max) energy consumption under guaranteed bit error rates.\par
Most of the UAV trajectory design schemes, e.g., \cite{ref4,ref9}, only apply a simple attenuation model that dominated by the distance. Some like \cite{ref12,ref13,ref19} discuss the difference between LoS and NLoS. These probabilistic models are easy and not considering the impact of buildings. To utilize the location-dependent and spatially varying channel as well as interference over the whole space, \cite{ref1} and \cite{ref21} gridded the space and study the parameters within each grid to construct a map to predict SINR and connectivity quality in the sampling locations separately. Centered on base stations, they have different coverage area with the building distributions. According to that, the authors of \cite{ref22} divided the coverage area into some sections and find trajectory in them. To get a more detailed information, \cite{ref21} utilized standard machine learning techniques such as k-nearest neighbors (KNN) or feedforward fully-connected artificial neural network (ANN) to train a complete CKM based on the finite measurements, to predict the CSI of unvisited locations. 
\par
Satellite positioning error is a frequently occurring factor caused by shadowing, mobility and time difference. Through experiments we recognized that position error significantly diminishes algorithm performance, an aspect neglected in existing models.
Inspired by the advancements of CKM, we aim to apply these new techniques to enhance the accuracy of our policy in real-world scenarios. Specifically, experiments will be conducted to construct a CKM capable of compensating for positioning errors. This CKM will be rigorously evaluated in the context of UAV communication and trajectory design, ensuring it effectively corrects for positioning inaccuracies and optimizes UAV operations. Our goal is to develop a robust solution that not only addresses the limitations of current models but also improves the reliability and efficiency of UAV deployments in complex environments.
\par
To utilize detailed and accurate channel information in time, experiments were performed in simulating software and redundant data were gathered to train a CKM which is robust to positioning errors to some extent. To verify the performace of this CKM, this article proposed a UAV trajectory design method called Positioning Error Correction(PEC)-PPO for flight time minimization in an emergency communication scenario with UAV as an aerial base station. The UAV was employed to send information to GUs, with the positions acquired from satellites or other sources. Then this CKM was applied in the formulated optimization problem to design the UAV communication under the constraints of UAV hardware limitations, transmit power, communication threshold and payload. Our contributions are summarized as follows.
\begin{itemize}
\item A robust CKM that is specifically designed to handle and compensate for positioning errors was trained. This advanced CKM leverages machine learning techniques to accurately predict the channel gain, even when user positioning information is imprecise.
\item We investigated an emergency communication scenario in which a UAV acts as a base station and establish a flight time minimization problem under communication constraints and physical restrictions. For this non-convex problem, it was transformed into a Markov decision process and the PPO-based method was used to solve it. In this process, the generated CKM was utilized, which has good noise resistance and a trajectory was successfully designed.
\item Experiments show that the greater the satellite positioning error, the longer the algorithm takes to finish the trajectory, and the lower the evaluation success rate. After using the generated CKM , the algorithm has a certain resistance to positioning errors. The flight trajectory no longer tends to wander due to errors, and the algorithm performance has also been improved.
\item Finally, in order to demonstrate the application characteristics of CKM, we verified its online update capability. Simulation with some dynamic changes of the environment was performed and data were re-collected for incremental learning to demonstrate the dynamic update capability of CKM.
\end{itemize}
\par
The remainder of this article is organized as follows. In Section II, the UAV-assisted emergency communication model was presented. Section III-A elaborated the explaination and the detailed construction method of positioning error resistant CKM. In Section III-B, a communication optimization problem was formulated. Then in Section III-C, the MDP formulation method was briefly introduced and the PEC-PPO algorithm was described in detail. Simulation results were stated in Section IV to evaluate the perfomance of the proposed algorithm. Finally, the conclusion is concluded in Section V.

\section{System Model}
We investigated a UAV-assisted communication system that an UAV serving as aerial base station is dispatched to fly in the space to transmit messages to ground users(GUs). In this system, satellites provide the positions of the GUs and feedback it to the UAV.
\par
\begin{figure}[t]
    \centering
    \includegraphics[width=\linewidth]{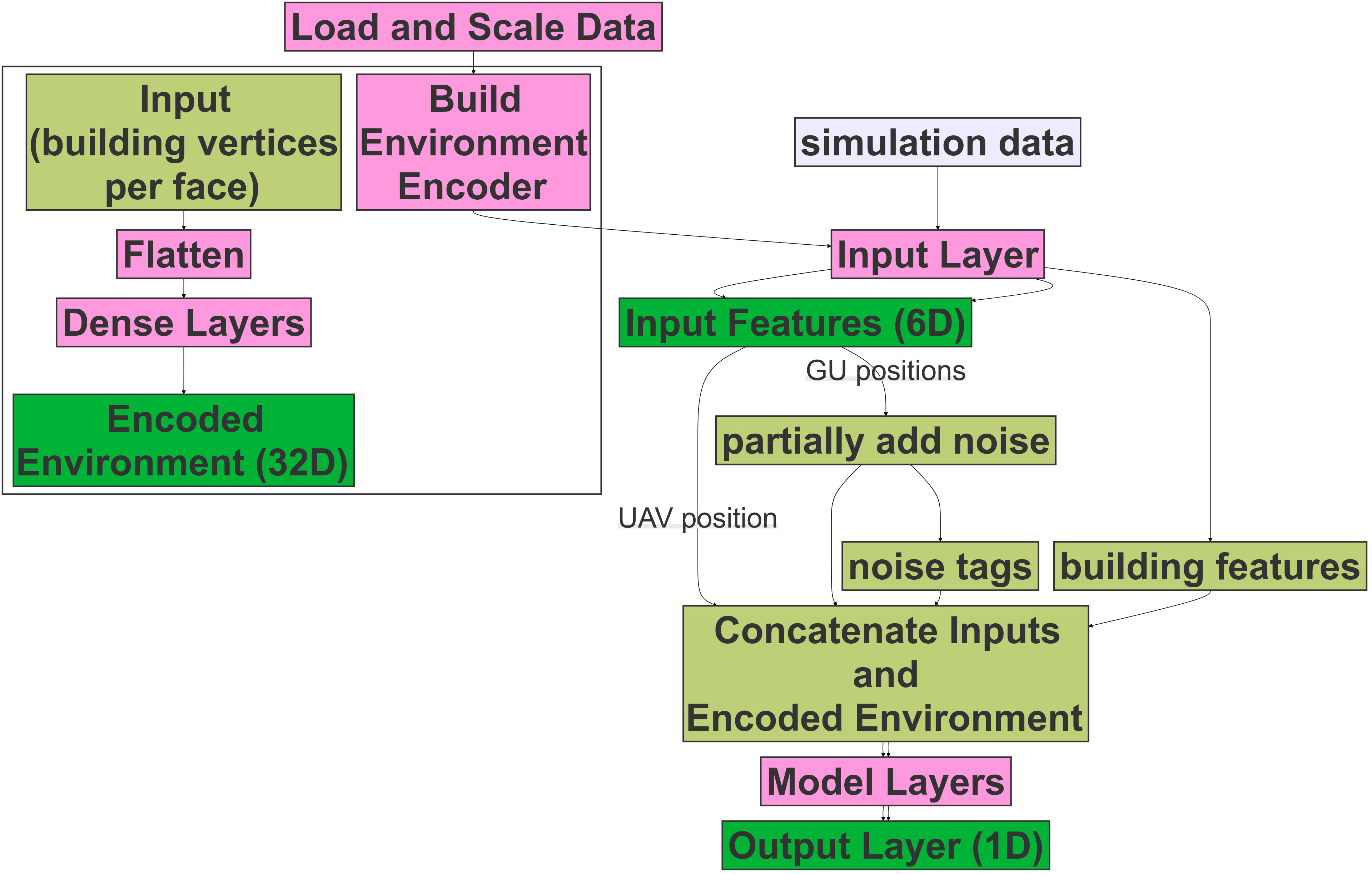}
    \caption{Process of  CKM construction. input: environment features, position pairs of GUs and the UAV in the simulation environment, noise labels, and output: channel gains}
    \label{fig4}
  \end{figure}
Our system is a cuboid with buildings, and the GUs are randomly distributed under a specified height. To avoid collision, the minimal height of UAV is set to $H_{min}$. We assumed that the GUs employ just one omnidirectional antennas while the UAV can work in a frequency division multiple access scheme to communicate with several GUs with each bandwidth $B$. The 3D Cartesian coordinates of the UAV and $i$-th GU is denoted by $\textbf{q}[t]=(x[t],y[t],z[t])$ and $\textbf{q}_i=(x_i,y_i,z_i)$ separately. To recall and charge the UAV in time, we set the starting position and end position of the UAV to be fixed at (0, 0, $H_{min}$). Here, $0 \leq t \leq T_{max}$. The states of the whole system keep unchanged during one time slot so the update of UAV's location between any two time slots can be derived,
\begin{equation}\label{1}
    \textbf{q}[t+1] = \textbf{q}[t] + \left[ 
    \begin{array}{ccc}
    \operatorname{cos}(\theta[t])\operatorname{cos}(\phi[t]) \\
    \operatorname{sin}(\theta[t])\operatorname{cos}(\phi[t]) \\
    \operatorname{sin}(\phi[t])
    \end{array} 
    \right ] \times V[t],
\end{equation}

\begin{equation}\label{2}
    V[t+1] = V[t]+a[t].
\end{equation}
In equation (\ref{1}), $\theta[t]$ and $\phi[t]$ are the direction angle and the elevation angle of UAV at time $t$.
\par
The simulation environment  is illustrated in Fig.\ref{fig2}. GUs are randomly placed and the UAV is allowed fly to operate for sufficient epochs to gather data, and then a network was trained to learn the mapping from UAV-GU position pairs to the channel gain.
Specifically, the CKM can be expressed as $\mathcal{M}:(\textbf{q}[t], \textbf{q}_i, \textbf{E}(t)) \rightarrow{} \mathcal{PL}[t]_i$. During the process of trajectory design, if the inputs are the position of UAV and $i$-th GU, and the concatenation with the environment features, the CKM will output the channel gain for throughput calculation.

\par
Based on that, the received signal from $i$-th GU can be expressed as
\begin{equation}
    P_R^i[t] = P^T[t] + \mathcal{PL}_i[t],
\end{equation}

and the corresponding data rate $R_i[t]$ is
\begin{equation}
    R_i[t] = Blog(1+\frac{P_i^R[t]}{\sigma^2}),
\end{equation}
where $P^T[t]$ is the transmit power allocated to UAV in time slot $t$ and $\sigma^2$ is the noise power in the channel.

\section{CKM-based PEC-PPO Scheme}
This section addresses the flight time minimization of the UAV-assisted communication network. A CKM is trained to correct most of the noises caused by positioning error, and then utilized in the formulation of the optimization problem. Then this high-dimensional, time-varying and non-convex problem is transformed into a Markov Decision Process. It's afterwards solved by a PPO-based algorithm and furthermore methods of online updating and power re-allocation are proposed to explore the superiority of the proposed CKM.
\subsection{Construction of CKM}
Considering that they might arise from various sources such as multipath, delay, and clock errors, positioning errors are inevitable in real-world applications. Once the UAV gets wrong locations of GUs, it would fail in right signal attenuation acquisition and GU selection, causing waste of communication time and quality. So it's crucial to construct a CKM that could correct most of the errors to maintain a good performance in trajectory design algorithm.
\begin{figure*}[t]
    \centering
    \includegraphics[width=\linewidth]{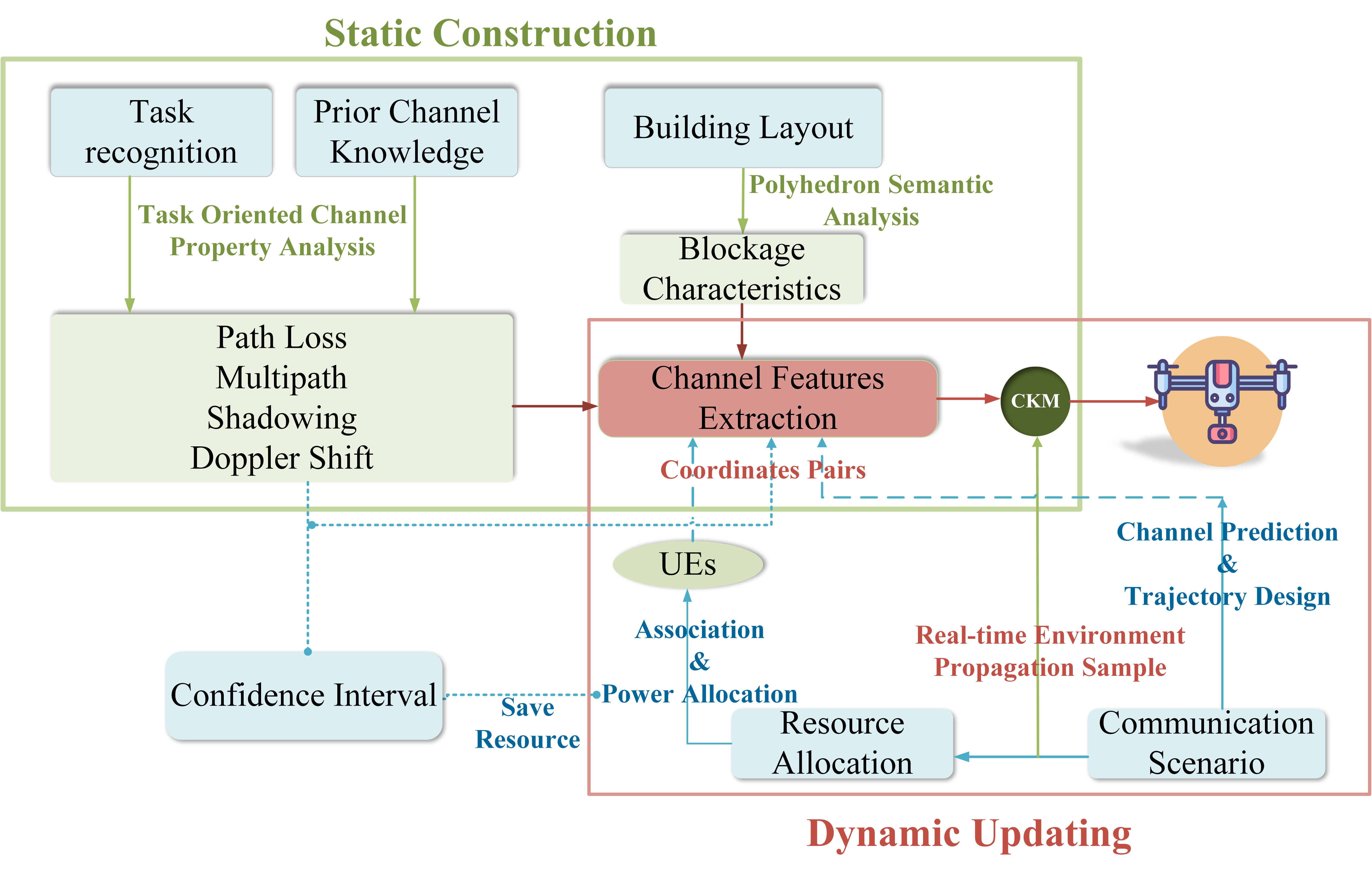}
    \caption{Detailed Construction and Utilization of CKM}
    \label{fig5}
  \end{figure*}
\par
CKM is a database that reflects the intrinsic wireless channel properties\cite{ref23}, offering any location-specific channel information to enhance environment-awareness. In our scenario, it can be regarded as a mapping from location pairs and environment features to the corresponding channel gains $\mathcal{M}:(\textbf{q}[t], \textbf{q}_i, \textbf{E}(t)) \rightarrow{} \mathcal{PL}[t]_i$ of the local environment.
\par
Circular Error Probability (CEP) is a statistical measure used to quantify the accuracy of GPS. It is defined as the radius of a circle within which $50\%$ of the measurements are expected to fall. In a spherical context, if the CEP is given as 5 meters, the true position of the points will follow a normal distribution with the center of the sphere as the mean and $\sigma=\frac{CEP}{0.6745}$. This relationship could be used in modeling the spatial distribution of positioning errors and integration in the CKM.
\par
To build the CKM, an environment was simulated in Wireless Insite, placing a transmitter trajectory to collect data, including the position pairs of GUs and the UAV, denoted by $\mathcal{F}_{pos}$, along with the corresponding channel gain $\mathcal{PL}$. As illustrated in Fig.\ref{fig4}, this article load the raw data and perform normalization to facilitate better learning and generalization while avoiding gradient explosion. This preprocessing step ensures that the data is in a suitable format for training the neural network.

\par
The vertices of buildings and plants are extracted as environment features $\mathcal{F}_{env}$ and then encoded and flattened into one dimensional. The concatenation of $\mathcal{F}_{env}$ and $\mathcal{F}_{pos}$ forms a 7-dimensional input to the network, with $\mathcal{PL}$ serving as the label. To simulate real-world scenarios where inputs are often noisy, noises are added to half of the input samples.
\par
To enable the neural network to distinguish between noisy and clean inputs, this article labels them differently and include these labels as additional input features. The network processes the inputs through multiple residual blocks, consisting of Dense layers with ReLU activations, Batch Normalization, and shortcut connections. These architectural components help the network learn complex mappings while maintaining gradient flow, enhancing the model's ability to generalize from the training data.
The final output layer of the network produces a single value prediction for the channel gain. Above approach ensures that the CKM can provide reliable, environment-aware channel information even in the presence of positioning errors.
\par
By integrating these advanced techniques, a CKM that not only captures the intrinsic wireless channel properties but also effectively corrects for positioning errors is constructed.

\subsection{Formulated Problem}
Denote the max flying time by $T_{max}$, and the solved finish time by $t_{end}$.

\par
Considering the hardware limitations that the UAV may encounter in practical scenarios, there should be several constraints. We denote the directional angle of UAV by $\theta[t]$, which refers to the angle between a reference direction and the line from the UAV to the original point of the system, measured in the horizontal plane. And the elevation angle is denoted by $\phi[t]$, which is the angle between the horizontal plane and the line of sight from UAV to the original point. Here the original point means the location of the UAV at the last time slot. 

Therefore, $\theta[t]$ and $\phi[t]$ should be limited to an interval,
\begin{equation}
    0 \leq \theta[t] \leq 2\pi,
\end{equation}
\begin{equation}
    -\frac{\pi}{2} \leq \phi[t] \leq \frac{\pi}{2}.
\end{equation}  
\par
Taking speed constraints into consideration, we have
\begin{equation}
    -a^{max}\leq a[t] \leq a^{max},
\end{equation}
and 
\begin{equation}
    0\leq V[t] \leq V_{max}.
\end{equation}
\par
To save energy, we set a max transmit power $P_{max}$ and the UAV won't always work in max power, so
\begin{equation}
    0\leq P^T[t] \leq P_{max}.
\end{equation}
\par
We set a predefined threshold $P_{min}$ for GUs, where the $i$-th GU could stay silent if and only if the received power $P_i^R[t] \leq P_{min}$, otherwise it will communicate with UAV. Furthermore, a binary association factor $\alpha_i[t]$ is defined to indicate whether the $i$-th GU is communicating with UAV, which can be written as
\begin{equation}  
    \alpha_i[t] = \left\{
        \begin{aligned}
        1, & \quad \text{if } P_i^R[t] \geq P_{min}, \\
        0, & \quad \text{otherwise}.
        \end{aligned}
        \right.
\end{equation} 
The following functions can be derived,
\begin{equation}
    \alpha_i[t](P_i^R[t]-P_{min}) \geq 0,
\end{equation}
\begin{equation}
    \alpha_i[t] \in \{0,1\}.
\end{equation}   
\par
Since the UAV is dispatched to serve as a base station, it's responsible to complete communication task. we set the payload for each GU is $\eta_{min}$ so the total throughput for each GU should meet this demand.
\begin{equation}
    \Sigma_{t=0}^{t=t_{end}}\alpha_i[t]R_i[t] \geq \eta_{min}.
\end{equation}
\par
The completing criteria is that the time reaches $T_{max}$ or the payload of each GU is cleared and the UAV flies back to the starting point. So our objective is to minimize the total time from the UAV starting to communicate with UEs to the UAV finishing the communication task and coming back. 
\par
The joint optimizing variables are $\alpha_i[t]$, $\textbf{q}[t]$ and $P_i^T[t]$, among them $\textbf{q}[t]$ and $P_i^T[t]$ are continuous while $\alpha_i[t]$ is binary.
The problem can be mathematically formulated as
\begin{small}
\begin{align} \label{13}
&\min_{\alpha_i[t],\textbf{q}[t],P_i^T[t]} \quad t_{end} \\
\text{s.t.}\quad
&t_{end} \leq t_{max} \tag{\ref{13}a}\\
&\textbf q[0] = \textbf q[t_{end}]=(0,0,H_{min}) \tag{\ref{13}b}\\
&0\leq \theta[t] \leq 2\pi,{\forall} t=0,1,...,t_{end} \tag{\ref{13}c}\\
&-\frac{\pi}{2}\leq \phi[t] \leq \frac{\pi}{2},{\forall} t=0,1,...,t_{end} \tag{\ref{13}d}\\
&0\leq V[t] \leq V_{max},{\forall} t=0,1,...,t_{end} \tag{\ref{13}e}\\
&-a^{max}\leq a[t] \leq a^{max},{\forall} t=0,1,...,t_{end} \tag{\ref{13}f}\\
&0\leq P^T[t] \leq P_{max},{\forall} t=0,1,...,t_{end},{\forall} i \in \mathcal I \tag{\ref{13}g}\\
&\alpha_i[t](P_i^R[t]-P_{min}) \geq 0,{\forall} t=0,1,...,t_{end},{\forall} i \in \mathcal I \tag{\ref{13}h}\\
&\alpha_i[t] \in \{0,1\},{\forall} t=0,1,...,t_{end},{\forall} i \in \mathcal I \tag{\ref{13}i}\\
&\Sigma_{t=0}^{t=t_{end}}\alpha_i[t]R_i[t] \geq \eta_{min},{\forall} i\in \mathcal I \tag{\ref{13}j}
\end{align}
\end{small}

\par
Firstly, to be recharged and assigned another mission, the UAV should fly to the original place when completing the task. In (\ref{13}b) we start at time 0 following the grammatical conventions of simulation.
\par
Above problem is high-dimensional(and the dimension would increase with the number of the GUs), time-varying and non-convex(which could be easily seen in (\ref{13}j)), it's complicated to apply SCA to make convex approximations, particularly that the variable of height is coupled in the denominator part. In contrast, DRL method can ignore the non-convexity and tackle it by interacting with the MDP environment.
\par

\subsection{MDP Formulation}
Combining the formulated optimization problem in Section III-B and the DRL framework, this article first translate the original problem as MDP structure and create an environment for the agent.
During the training process of PPO2, the initial state information is inputted into the actor network, which outputs the distribution parameters of the action. These parameters are then sampled to get an action, and the reward function and next state are calculated. This information is stored for several cycles without updating to gather a batch of experience. After collecting sufficient data, the final state $s'$ is inputted into the critic network  to be valued. The advantage function and loss are then computed based on the collected data. The advantage function is calculated to measure how much better or worse an action is compared to the expected value of the state. Using the computed advantages, the parameters of the critic-NN are updated by minimizing the value function loss. Next, the parameters in the actor network  are updated. PPO2 alternates between collecting experience through interaction with the environment and updating the actor and critic networks based on the collected data. This process ensures stable and efficient learning, allowing the agent to effectively improve its performance over time.
\par
The state needs to be reward-centric and contain the required observations, while the action needs to ensure the complete characterization of the state change.
\subsubsection{State Space}
The state space promptly feeds back the information that needs to be observed during the training process. In order to avoid overfitting caused by useless information, it must only contain the parameters of interest. In the simulation process, the state information is defined as a one-dimensional vector that contains the specified variables at all times. State $s_t$ in the time slot $t$, $\forall t$ includes:
\begin{enumerate}[label=\alph*)]
    \item $(x[t],y[t],z[t])$: The normalized 3-D coordinate of the UAV.
    \item $V[t]$: The normalized velocity of the UAV.
    \item $(\theta[t],\phi[t])$: The normalized direction and elevation angle of the UAV.
    \item $(x_i,y_i,z_i)$: The normalized 3-D coordinate of the $i$-th GU.
    \item $\alpha_i[t] \in [0,1]$: The communication indicator of the $i$-th GU and UAV at time slot $t$. If $\alpha_i[t]=1$, the received power of $i$-th GU is over the threshold and it is communicating with the UAV.
    \item $\eta_i[t] \in [0,1]$: The remaining payload of the $i$-th GU. When it turns into 0, the $i$-th GU doesn't need to communicate with UAV. This factor could help to observe the detailed communication state in each time and determine whether the communication task is finished.
    \item $d_i[t] \in [0,1]$: The distance between the UAV and the $i$-th GU.
\end{enumerate}
In order to avoid overfitting, ensure the generalization of the algorithm, and prevent the algorithm from being too biased towards certain parameters with large values, the above values are normalized.
\subsubsection{Action Space}
The design of the action space needs to ensure that the scale is consistent with the state space. For example, in this paper, the distance and coordinate units of the action space and state space are meters. The action space contains all the variables that the drone needs to update its actions, including the following parts:
\begin{enumerate}[label=\alph*)]
    \item $a[t] \in [-1,1]$: The acceleration of the UAV. This parameter can be used to calculate the speed of the drone and then calculate the coordinates.
    \item $\theta[t] \in [0,1]$: The total range is $2\pi$. 
    \item $\phi[t] \in [-1,1]$: The total range is $\frac{\pi}{2}$.
    \item $P^T[t]$: The transmit power of the UAV. It does not always maintain the maximum power value, which can save energy.
\end{enumerate}
When interacting with the environment, the UAV operates by selecting actions randomly from this action space and then continuously collects and records state information.By randomly exploring its action space and systematically gathering state information, the UAV builds a comprehensive understanding of its operational environment. This exploratory process is essential for developing robust strategies that can adapt to varying conditions and unforeseen challenges, ultimately enhancing the UAV's efficiency and effectiveness in achieving its goals.  
\subsubsection{Reward Function}
In order to minimize the flight time, the reward function needs to be negatively correlated with time.
Completing the communication mission and flying back to the original location is the core goal, but this reward starts too late, which will lead to slow convergence, so it is necessary to add some sub-goals to guide the UAV to explore in right trends.
\par
First of all, the range of movement of the UAV needs to be limited to the space, so every time it's needed to check whether it exceeds the boundary, and if it exceeds the boundary, a penalty is imposed.
\begin{equation}
    R_1 = -r_1\times punishment,
\end{equation}
\par
Whenever a new GU has completed a communication task, a certain reward is given. In order to reduce the time to complete the task, time weighting is used.
\begin{equation}
    R_2 = r_2 \times N_{new} \times (T_{max} - t),
\end{equation}
$N_{new}$ is the number of GUs that just completed the communication task, which can be calculated from $\eta_i[t]$ in state space.
When the UAV fished communicate with all GUs and flies back to the original point, $R_3$ works,
\begin{equation}
    R_3 = r_3 \times (T_{max} - t).
\end{equation}
The sum of the three is used as the reward function for model training. If the model can be trained successfully, the reward function will first increase and then converge to a stable value. Correspondingly, the time to complete the task will gradually decrease and converge to a stable value.
\subsection{PPO2-based UAV Trajectory Design with CKM}
To solve the trajectory design problem, the PPO method and the proposed CKM are combined together. The pseudocode is listed in Algorithm1. We use unshared actor and critic networks to ensure that the updates of the two networks do not interfere with each other. In order to improve the generalization ability of the algorithm in various situations, the GUs are generated randomly. \textit{adaptive RL-env, noise correction, and online scheduling} techniques are applied to make the algorithm noise-robust and environment-aware.
\begin{algorithm}[t]
\caption{PEC-PPO}\label{Algorithm1}
\begin{algorithmic}
\STATE Initialize actor network parameters $\theta^0$ and critic network parameters $\phi^0$
    \STATE Train CKM using collected data
    \FOR{episode = 1, \dots, M}
        \STATE Reset the environment and get an initial state $s_1$
        \FOR{step = 0, 1, \dots, $T_{max}$}
            \STATE Take action $a_t = \pi_{\theta}(s_t)$
            \IF{the action is out of boundary}
                \STATE Clip the action following constraints
                \STATE $punishment   += 1$
            \ENDIF
            \STATE $\textbf{q}_{t+1} \leftarrow{} a_t,\textbf{q}_t$
            \STATE Use CKM to get channel gain at the new position with error:
            \STATE \quad add($\textbf{q}_{t+1}$, CEP), flag=1 $\rightarrow{} \mathcal{H}_n$
            \STATE \quad Obtain channel gain 
            \STATE Select GUs index $\alpha_i$
            \STATE Adjust transmission power based on predicted value accuracy
            \STATE Use CKM to get error-free channel gain for rate calculation:
            \STATE \quad $\textbf{q}_{t+1}$, flag=0 $\rightarrow{} \mathcal{H}_o$
            \STATE \quad Calculate rate based on obtained channel gain
            \STATE Collect $(s_t, a_t, r_t)$
            \STATE Update the state $s_{t+1}$
            \IF{step \% L == 0}
                \STATE Compute advantage estimates using Critic network
                \STATE Update $\theta$ by gradient method
                \STATE Update $\phi$ by gradient method
            \ENDIF
            \IF{terminated}
                \STATE break   
            \ENDIF
        \ENDFOR
    \ENDFOR
\end{algorithmic}
\label{alg1}
\end{algorithm}
\par
1) \textit{adaptive RL-env}.
We incorporated the CKM model into the environment, leveraging it for user association and rate computation. During this process, the physical environment influences the UAV's decision-making, and changes in the UAV's actions subsequently alter the CKM input, ultimately affecting the calculated values. This interactive process facilitates a dynamic response to the communication environment. By integrating the CKM model, the environment becomes more realistic and adaptive, allowing the reinforcement learning agent to learn robust strategies that are effective in a wide range of unpredictable real-world scenarios.
\par
2) \textit{noise correction}.
The trained CKM is capable of predicting channel gains for both accurate and erroneous positions. In the UAV's interaction with the environment, during the user selection phase, CEP noise is added to the position to represent potential errors. The erroneous position, along with a tag indicating $flag=1$, is inputted into the CKM to obtain the corresponding channel gain, which is then used to determine whether to select a user for communication. During the communication phase, the original position and a tag indicating $flag=0$ are inputted into the CKM to calculate the communication rate based on the predicted channel gain.These noise correction strategies ensure that the CKM can effectively mitigate the impact of positional inaccuracies, thereby enhancing the UAV's ability to maintain reliable communication under varying conditions and achieving significant noise resistance.
\par
3) \textit{online scheduling}.
To enhance the dynamism of our communication strategy, we implement an online scheduling approach that adjusts power selection based on the channel gain difference between noisy and noise-free positions. This technique supports decision-making by leveraging the real-time accuracy of the CKM. Specifically, when the channel gain difference is small and near the communication threshold, we increase the transmission power to ensure reliable communication. Conversely, when the gap is substantial and far from the threshold, we opt to abandon the communication attempt to conserve resources.
Additionally, we employ incremental learning to continuously train the model with new data as the environment changes. This enables the UAV to adapt its strategy online, maintaining good performance even in dynamic and unpredictable scenarios.

\par
Our algorithm flow is as follows: first, the collected data are used to train the CKM and then the reinforcement learning environment is initialized. In each training episode, the UAV's actions and states are initialized. Based on the current state, the actor network selects a flight action and calculates the UAV's next position. The CKM is then used to obtain the channel gain at the new position, which corresponds to the user's position with errors. This information is used to select the GUs' indexes for communication. The transmission power is adjusted according to the accuracy of the predicted value at this time. The CKM is used again to obtain the error-free channel gain for rate calculation. Subsequently, a series of flags judgments are performed, and the state is saved to determine whether the training should end. If not, the process proceeds to the next round. In the process of using CKM, if the environment changes, online training and updating will be performed.
\section{Simulation Results}
\begin{figure}[t]
    \centering
    \includegraphics[width=\linewidth]{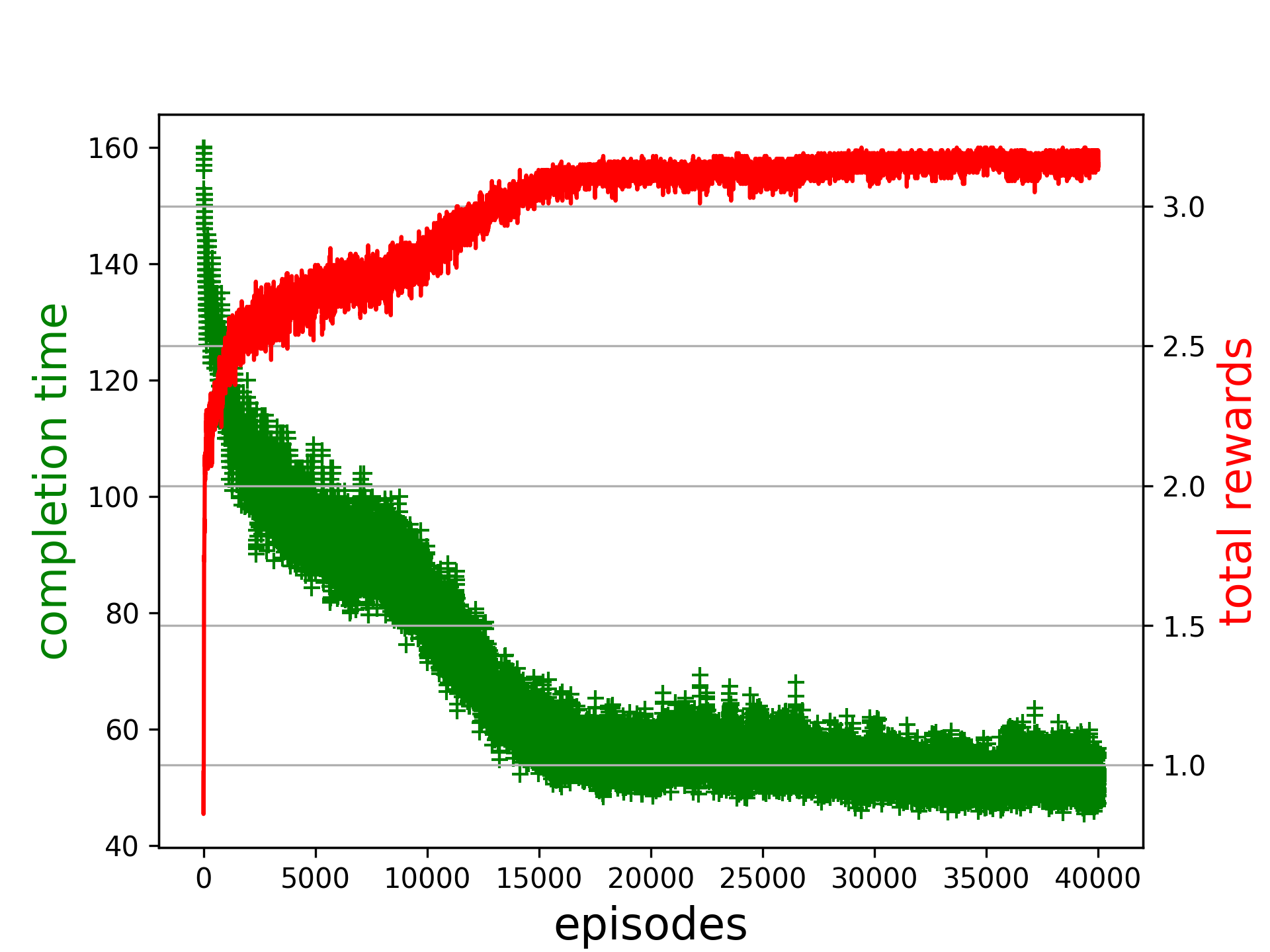}
    \caption{The convergence of completion time(green) and rewards(red) under max time $160s$ and reward scaling factors $r_1=0.000001,r_2=0.0012,r_3=0.005$}
    \label{fig9}
  \end{figure} 
\begin{figure}[t]
    \centering
    \includegraphics[width=\linewidth]{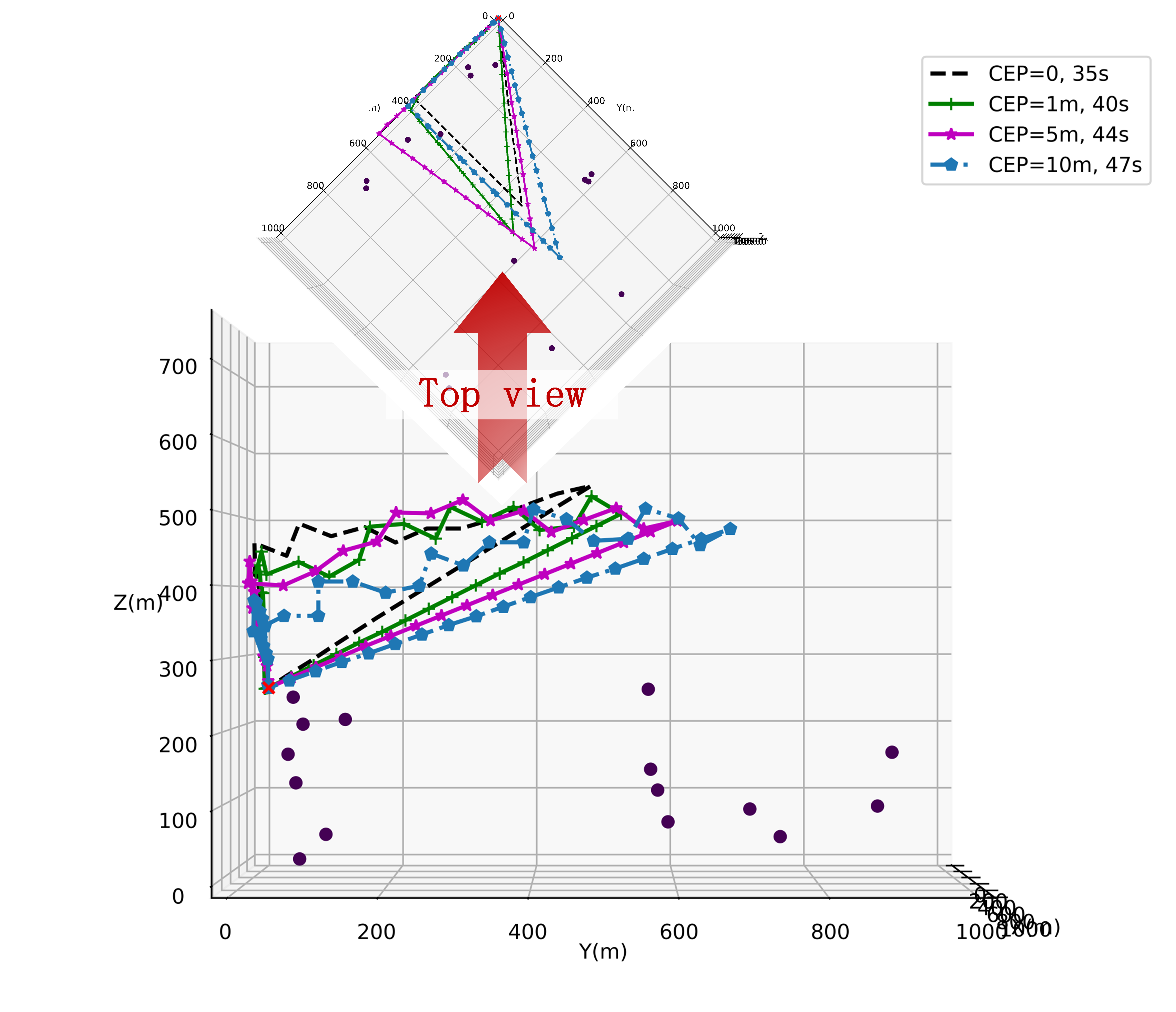}
    \caption{trajectory of CKM-PPO under the positioning error of CEP=0, 1, 5, 10m}
\label{fig6}
\end{figure} 
In this section, our simulation results of the proposed PEC-PPO algorithm is presented.
\subsection{Simulation Settings}
As shown in Fig.\ref{fig2}, our communication space is of size $1000m \times 1000m \times 750m$ and the UAV is allowed to fly between $250m$ and $750m$. The max transmit power of the UAV is $P_{max}=26dBm$, the communication threshold is $P_{min}=-70dBm$, and the noise power is $\sigma^2=-104dBm$. The information size for each GU is $26Mb$ and the bandwidth allocated to each of them is $B=1MHz$. The max flight velocity of the UAV is $50m/s$ and the max acceleration is $20m/s^2$. The communication frequency is $2GHz$. Part of the training parameters are listed in TABLE \ref{table:table1}
\begin{table}[h!]
\centering
\caption{Main Training Parameters}
\label{table:table1}
\begin{tabularx}{\linewidth}{@{}>{\raggedright\arraybackslash}X >{\raggedright\arraybackslash}X@{}}
\toprule
\textbf{Simulation parameter} & \textbf{Value} \\
\midrule
architecture of actor-NN & $(96, 256, 256, 4)$ \\
architecture of critic-NN & $(96, 256, 256, 1)$ \\
Maximum number of episodes & $M = 40000$ \\
Maximum number of steps in each episode & $T_{max} = 160$ \\
Discount factor & $\gamma = 0.99$ \\
Learning rate & $l_r = 1e-5$ \\
Clip range & $cr = 0.2$ \\
\bottomrule
\end{tabularx}
\end{table}
\par
In order to evaluate the performance of the algorithm, several algorithms are used for comparison with the proposed PEC-PPO algorithm.
\par
1) \textit{LoS-based PPO}: The UAV uses LoS probability model to calculate the signal attenuation between the user and the user according to $P_{los}(r_i[t],s_i[t])=\frac{1}{1+ae^{-b(\tan^{-1}\frac{h_i[t]}{r_i[t]}-a)}}$ and $L_{los/nlos}(d_i[t]) = 20log(\frac{4\pi f_cd_i[t]}{c}) + \epsilon_{los/nlos}$.
\par
2) \textit{CKM-PPO}: The UAV uses ordinary CKM to obtain signal attenuation and has no resistance to positioning noise.
\par
3) \textit{Offline Scheduling(OS)-PPO}: The UAV just applied the noise robust CKM offline and no update with the dynamic environment would be implemented.

\subsection{Results and Analysis}
\begin{figure*}[t]
    \centering
    \includegraphics[width=\linewidth]{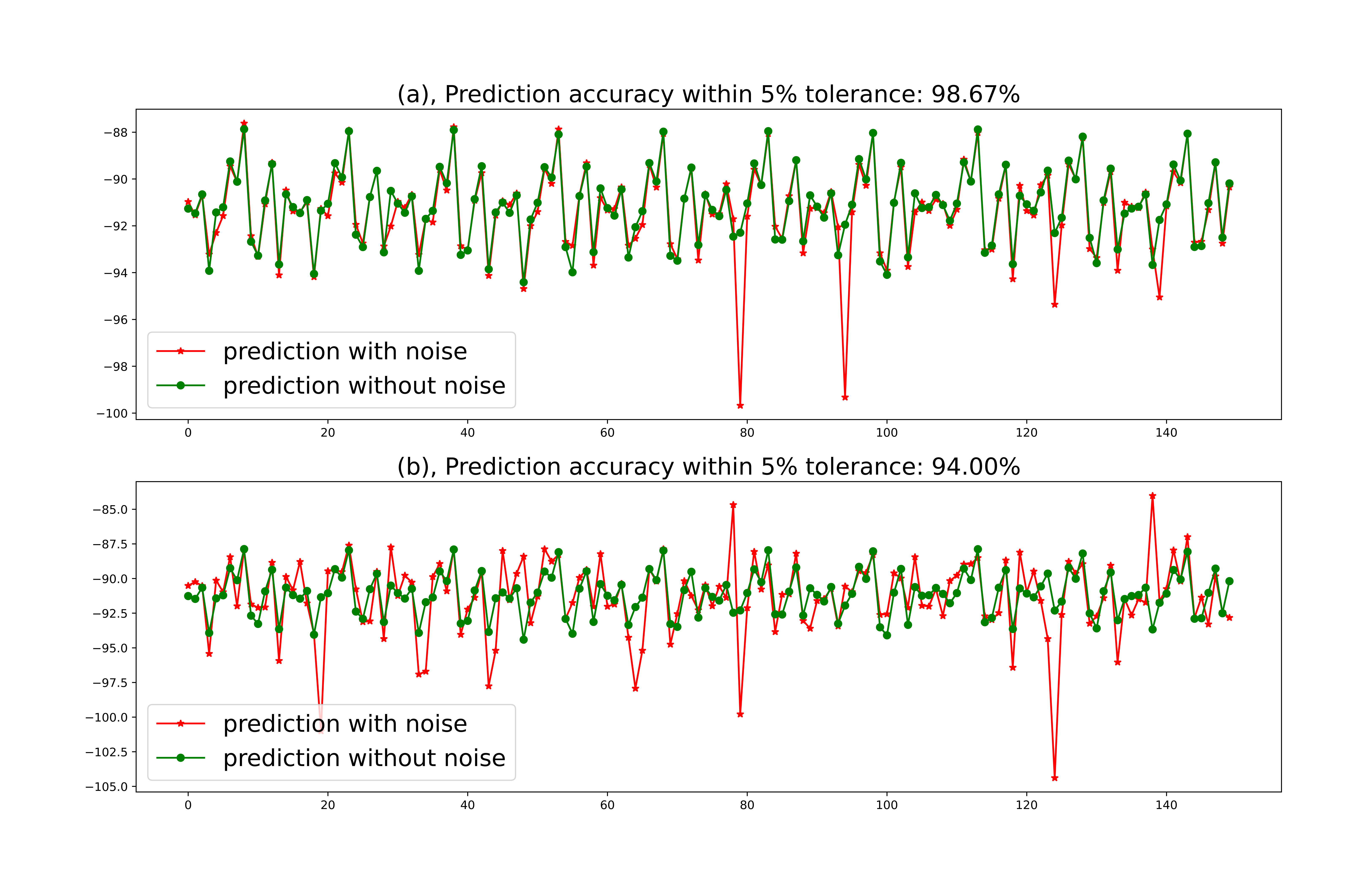}
    \caption{CKM predictions of channel gain for true and errored GUs' positions, (a) CEP=5m, (b) CEP=10m}
    \label{fig7}
  \end{figure*} 
During each training, the user's position are randomlu generated and the UAV start from a fixed original point. During the evaluation, we average the results of 500 times to display the data, and selecte the group with the shortest flight time for visualization and comparison.
\par
First, it should be checked whether the algorithm has converged. As shown in the Fig.\ref{fig9}, after 40000 episodes of training, the time it takes for the UAV to complete the task gradually converges to about $50s$, and the corresponding reward value gradually increases to 3. This shows that our algorithm has the ability to converge in the similar class of the proposed scenario.

\par

When the UAV flies high, the LoS probability could increase and then decrease, which means it could get a maximal coverage in one point unimodally. However, the higher it flies, the lower the user's receiving power is, so the UAV needs to make a compromise between coverage and receiving power. It can also be seen from Fig.\ref{fig6} that the UAV does not fly upward all the time, but completes the communication with the user when it reaches about 550 meters.

\par

As Fig.\ref{fig6} shows, this article adds different positioning errors to the GUs' position and apply the normal CKM to design trajectory. CEP will interfere with training and affect the optimality of task decision-making. During the mission execution phase, the UAV's flight trend is to fly upwards, but the existence of errors will lead to wrong decisions. Particularly, if the measured distance is closer than the actual distance, and the decision is to descend to communicate, while the actual received signal strength is small, the communication will fail and need to establish a connection again. If the measured distance is farther than the actual distance, and the decision is not to establish a connection and continue to fly upward to increase the coverage area, thereby wasting communication opportunities, the flight trajectory will show as up and down twists and turns. The larger error forces the UAV to fly deeper to establish a connection with users in the corner. At the same time, due to the incorrect estimation of the received signal strength, the trajectory fluctuates more and the overall flight altitude is lower to reduce path loss. For small CEP, the UAV will sacrifice part of the coverage range to obtain a stronger received signal. For large CEP, the compromise behavior at this time will be more inclined to a larger coverage range to search for more users who reach the communication threshold, and the high error rate causes some users to complete communication in the return stage.
\begin{figure*}[t]
    \centering
    \includegraphics[width=\textwidth]{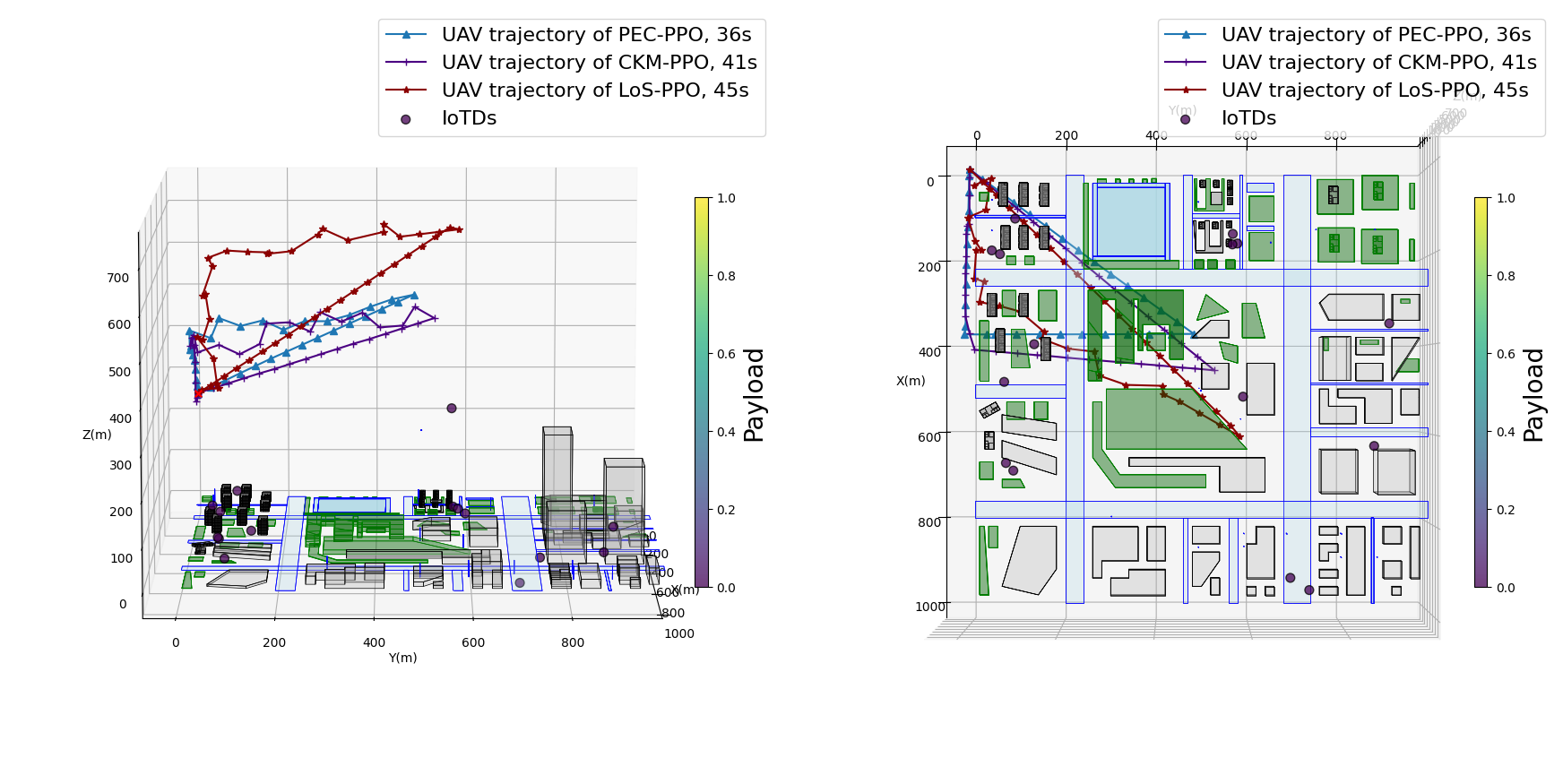}
    \caption{The trajectory of PEC-PPO, CKM-PPO and LoS-PPO under CEP = 5m}
    \label{fig77}
\end{figure*}
\begin{figure*}[t]
    \centering
    \includegraphics[width=\textwidth]{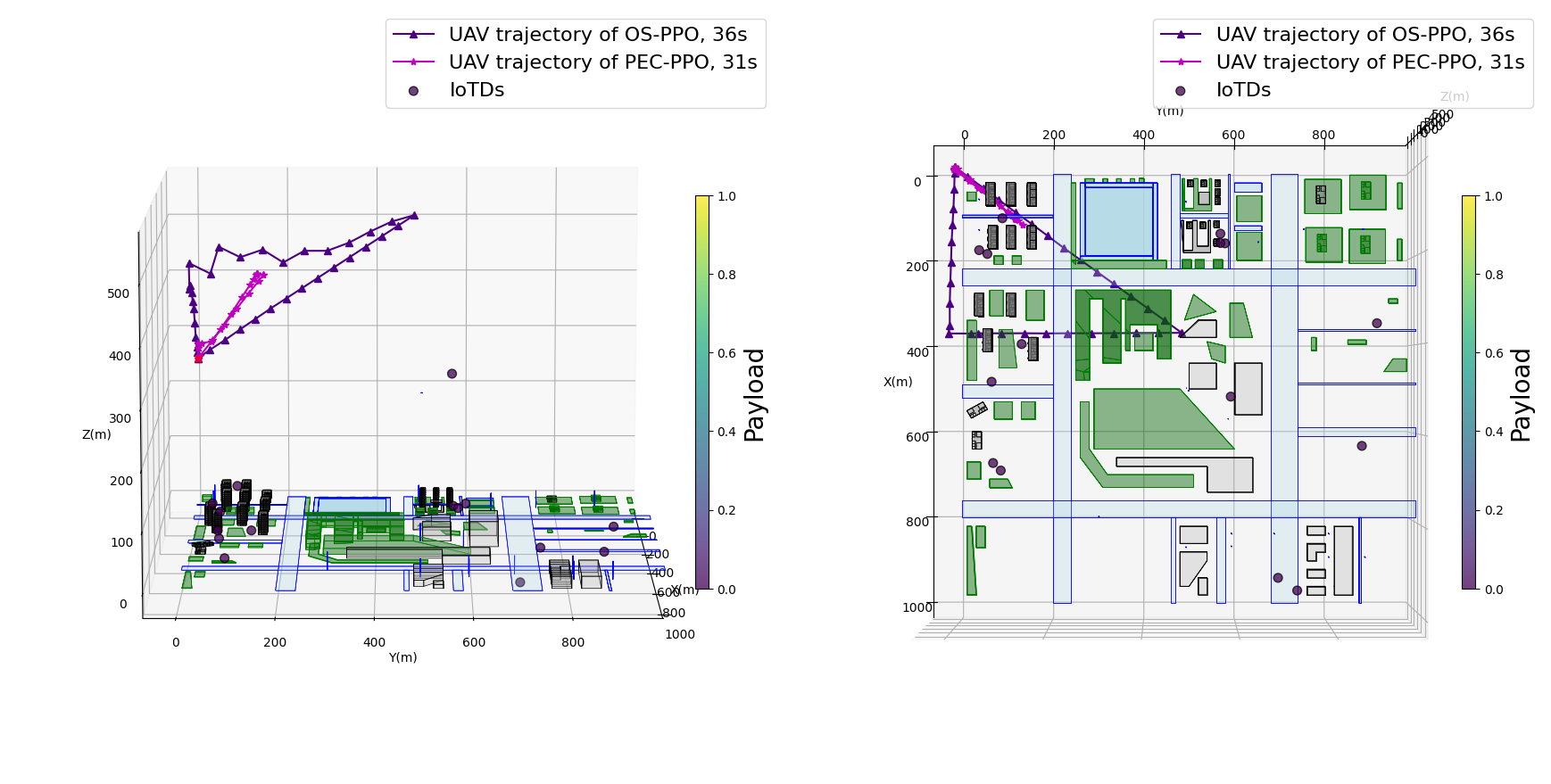}
    \caption{The trajectory of PEC-PPO and OS-PPO under CEP = 5m, with part of the tall buildings damaged}
    \label{fig8}
\end{figure*}
\par
Then a CKM is trained using noisy data with CEP=5 and its prediction performance for data with and without positioning errors is evaluated. It can be seen that the trained CKM is somewhat resistant to errors. It is also worth noting that even though the used data are errored with CEP=5m, the CKM is still robust to errors of other values. Fig.\ref{fig7} shows that for data with CEP=25m, our proposed CKM is still feasible for most of the data. It can also be seen that the predicted values of some individual points have large differences. When the received signal threshold is between the noisy and noise-free received powers, this will cause errors in the judgment results and affect communication. We will adjust the strategy online later.

\par
The trained CKM is incorporated into the trajectory design algorithm. Specifically, in the user selection phase, the drone calculates the coverage range based on the acquired user positions with errors and selects the communication user. In the rate calculation phase, the received power and throughput are calculated based on the actual user positions. Through the training results it can further be found the error resistance of the PEC-PPO algorithm.

\par
It can be found that the highest point of the LOS trajectory reaches more than 700 m, while the highest points of the two CKM-based algorithm trajectories are only more than 500 m. According to the relationship between coverage range and flight altitude analyzed above, the relative flight altitude corresponding to the maximum coverage range calculated based on the LOS channel is 550 m, and the highest user is 200m. Thus the UAV using LoS-PPO needs to fly at about 750 m to obtain the maximum coverage area while for CKM-based algorithm, good communication environment makes it possible to fly lower.
\par
As can be seen from Fig.\ref{fig77}, using the trained CKM for communication can effectively avoid CEP errors, which is manifested in that the behavior trajectory is no longer reciprocating. In addition, the LoS model uses a general probability distribution of buildings, but the buildings in our simulation are not so dense, so the channel conditions predicted by CKM are much better than that calculated by LOS. Therefore, the flight time obtained by the PEC-PPO and CKM-PPO algorithms is shorter, and the UAV completes the communication and turns back in a place that is not so deep. As for PEC-PPO, due to the addition of noise robustness training, part of the positioning errors can be corrected, so the algorithm performance is better than CKM-PPO. And by observing the dynamic trajectory, it is found that the UAV almost no longer has round-trip behavior caused by errors, which shows the effectiveness of our proposed algorithm in eliminating positioning errors.

\par
To demonstrate the algorithm’s adaptive capabilities, the algorithm is tested in two scenarios: building changes and automatic power adjustment, and compared it with OS-PPO, which does not have the ability to update.
\par
Data are re-collected in the new environment and incremental learning is applied to train CKM, and apply it to the drone trajectory design. This step can be added during the drone training process or re-trained. As can be seen from Fig.\ref{fig8}, since part of the high buildings are removed, the communication conditions have improved, and the UAV does not need to fly far to establish a connection with the GUs to complete the communication, so the trajectory range has become much narrower. The OS-PPO algorithm used for comparison does not have the ability of dynamic perception and cannot change according to the environment.
\par
Besides, if the confidence interval is wide and the median is a little far from the threshold, the communication would be gave up; if the confidence interval is narrow and close to the threshold, the communication would continue at max power. We make multiple predictions and calculate confidence intervals, get the median of the interval $I_{med}$, judge it and allocate power in the following way.

\begin{equation}  
    P^T[t] = \left\{
        \begin{aligned}
        & P_{max}, && \text{if } I_{med} \geq P_{max} - P_{min}, \\
        & 0, && \text{otherwise}.
        \end{aligned}
        \right. 
\end{equation} 

\par
\begin{figure}[t]
    \centering
    \includegraphics[width=\linewidth]{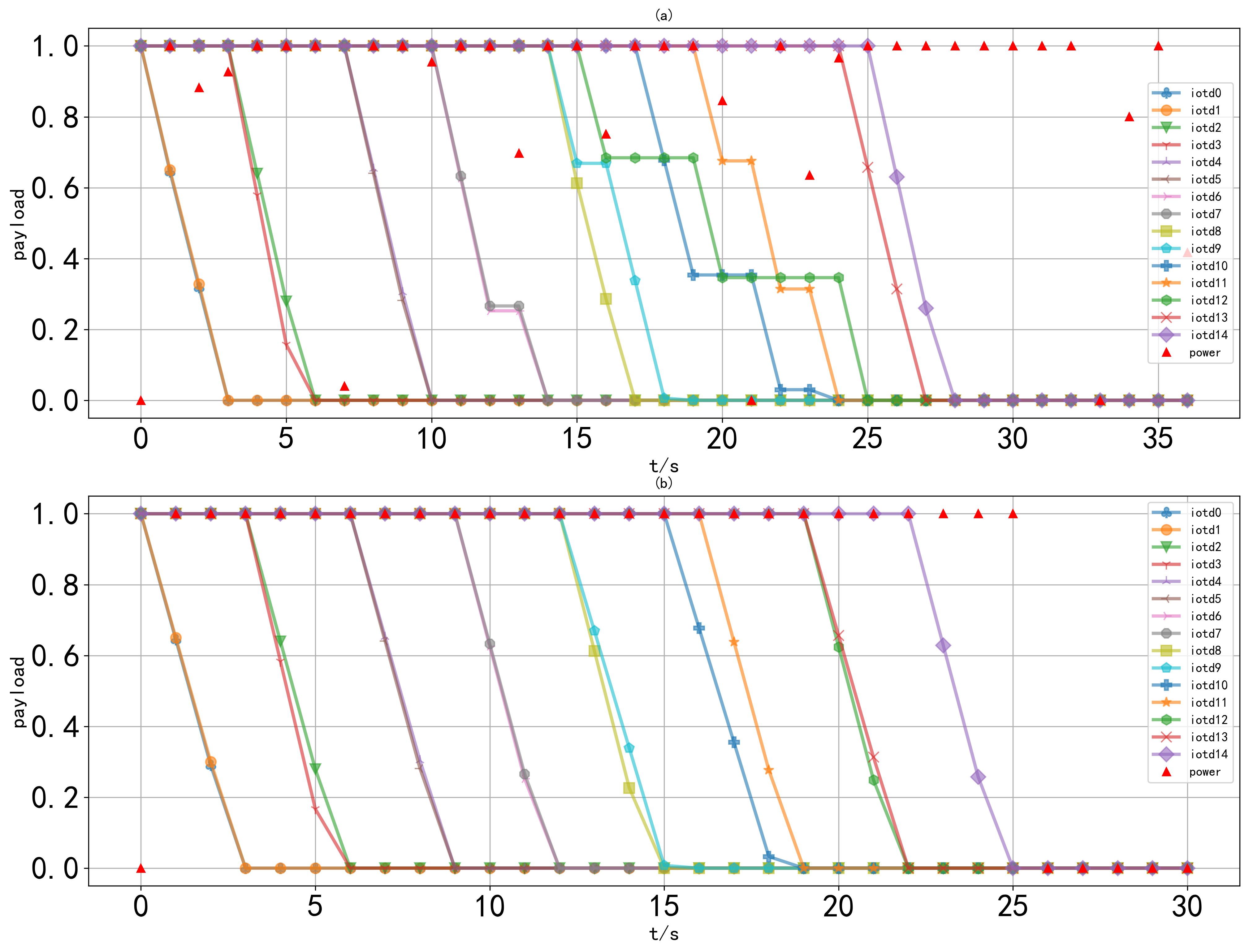}
    \caption{the transmit power and GU payload with time, (a) OS-PPO, (b) PEC-PPO}
    \label{fig10}
  \end{figure}

Experimental results show that this power allocation allows communication at maximum power when channel conditions are good and quickly clears user loads. When the CKM prediction credibility is not high enough, energy can be saved by abandoning communication. Fig.\ref*{fig10} shows that in the training of the UAV, since there is no power-related reward, the power is still not zero after the communication ends, which leads to waste. However, the power adjustment strategy can effectively utilize the power for communication, and finally achieve the goal of saving energy. In addition, adaptive power selection allows users to receive higher power, thereby eliminating the platform phenomenon in Fig.\ref*{fig10}(a) and ensuring that the UAV is well connected to the user at all times.

\section{Conclusion}
This article investigated an emergency communication scenario where a UAV acts as a base station, and addressed a flight time minimization problem under communication constraints by transforming it into a Markov decision process. A CKM was constructed with input features designed specially for positioning error robustness. The CKM was utilized with deep reinforcement learnign method in a dynamic environment. Simulation results showed that the CKM demonstrates strong noise resistance and effective trajectory planning, improving algorithm performance despite positioning errors. Additionally, the CKM showed dynamic update capability, adapting to changes in the communication environment.
In future work, we plan to incorporate more accurate channel models and simulation methodologies to devise refined correction strategies and enhance the precision and reliability of our system under various operational conditions.


%

\vspace{11pt}

\end{document}